# Optimum annular focusing by a phase plate

VICTOR ARRIZÓN, ULISES RUIZ, DILIA AGUIRRE-OLIVAS, AND GABRIEL MELLADO-VILLASEÑOR.

*Instituto Nacional de Astrofísica, Óptica y Electrónica, Puebla 72000, México*

**Conventional light focusing, i. e. concentration of an extended optical field within a small area around a point, is a frequently used process in Optics. An important extension to conventional focusing is the generation of the annular focal field of an optical beam. We discuss a simple optical setup that achieves this kind of focusing employing a phase plate as unique optical component. We first establish the class of beams that being transmitted through the phase plate can be focused into an annular field with topological charge of arbitrary integer order $q$. Then, for each beam in this class we determine the plate transmittance that generates the focal field with the maximum possible peak intensity. In particular, we discuss and implement experimentally the optimum annular focusing of a Gaussian beam. The attributes of optimum annular focal fields, namely the high peak intensity, intensity gradient and narrow annular section, are advantageous for different applications of such structured fields.**

Light focusing, one of the processes more used in optics, is usually realized by a lens. The perfect focal point, with an infinitely small size, cannot be achieved because of the finite extension of both the lens and the optical field to be focused [1]. Here we discuss focusing of monochromatic light into an annular focal field, assuming that it is modulated by an azimuthal phase of arbitrary integer order $q$. The inclusion of the topological charge transforms the focal field into an annular vortex (AV). This type of structured fields, are useful in several applications, e.g. optical trapping with orbital angular momentum transference [2-4], lithography [5, 6], high-resolution fluorescence microscopy [7], quantum entanglement [8-10], and vortex coronography [11, 12]. As occurs in conventional focusing, the generation of an infinitely narrow (or perfect) AV [13-16] is impossible. Therefore, it is important to establish the optimum approximation of this field that can be physically implemented. Here we consider that the optimum AV is the one with the maximum possible peak intensity. The maximum intensity in AVs implies other attributes, as a narrow transverse section and a high intensity gradient that may offer advantages in different applications of such fields.

The generation of an optimum AV at the Fourier domain of a phase diffractive element, which is illuminated by a Gaussian beam, has been recently reported [17]. Here we discuss the simplest method for annular focusing of an arbitrary beam. This method employs a phase plate as unique optical component, which modulates the complex amplitude of the beam. The AV is obtained, by free propagation of the modulated beam, at a specific distance from the plate. As first step, we establish the class of beams that can generate an AV using this simple method. Then, we determine the phase plate transmittance required to achieve the optimum annular focusing of the beams in this class.

One of the optical fields that belong to the referred class is the Gaussian beam (GB). We give special attention to the GB because it is efficiently generated, as the usual mode of a laser. In particular, we establish the explicit transmittance of the phase plate that generates the optimum annular focal field of a GB, and employ a phase spatial light modulator (SLM) for its generation.

To discuss annular focusing of a beam we refer to the setup depicted in Fig. 1. In this setup, the input beam (B) is passed through a phase plate (PP), and the AV is generated, by free propagation of the transmitted beam, at a distance $z$ from the plate.

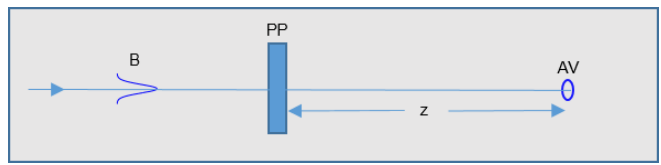

Fig. 1. Simple setup to generate the annular focusing of a beam (B).

The optical fields are expressed in polar coordinates $(\xi,\phi)$ at the plate plane and $(r,\theta)$ at the focal field plane. The complex amplitude of a generic AV, with integer topological charge $q$, is

$$h(r,\theta) = F(r)\exp(iq\theta), \qquad (1)$$

whose radial factor $F(r)$ is specified below. Considering that the AV, with the separable form in Eq. (1), is obtained by free propagation

of the field transmitted by the plate, it is easy to prove that this field must have the separable form

$$f(\xi,\phi) = a(\xi)\exp[i\beta(\xi)]\exp(iq\phi), \quad (2)$$

where the amplitude $a(\xi)$ is a non-negative function and $\beta(\xi)$ is a radial phase function to be determined. This result is obtained using either exact or paraxial scalar field propagation.

Now we assume that the complex amplitude of the beam that illuminates the phase plate is $g(\xi,\phi)=|g(\xi,\phi)|\exp[i\alpha(\xi,\phi)]$, with amplitude $|g(\xi,\phi)|$ and phase $\alpha(\xi,\phi)$. Thus, denoting the phase plate transmittance as $p(\xi,\phi)$, we establish the identity $f(\xi,\phi)=g(\xi,\phi)p(\xi,\phi)$. Considering the explicit terms in this equation it is easy to show that the complex amplitude of the input beam is given by

$$g(\xi,\phi) = a(\xi)\exp[i\alpha(\xi,\phi)], \quad (3)$$

with amplitude dependent only in the radial coordinate $\xi$ and arbitrary phase $\alpha(\xi,\phi)$; while the transmittance of the phase plate is given by

$$p(\xi,\phi) = \exp[i\beta(\xi)]\exp[-i\alpha(\xi,\phi)]\exp(iq\phi). \quad (4)$$

The unknown phase $\beta(\xi)$ in Eq. (4) is next specified in order to give desired attributes to the radial factor $F(r)$ of the AV field. Performing the Fresnel propagation of the field $f(\xi,\phi)$ [Eq. (2)] to a distance $z$ one obtains the field $h(r,\theta)$ [Eq. (1)], whose radial factor is (omitting constants)

$$F(r) = \exp\left(i\frac{kr^2}{2z}\right)\int_0^\infty \xi a(\xi)\exp[i\chi(\xi)]J_q\left(2\pi\frac{r}{\lambda z}\xi\right)d\xi, \quad (5)$$

where $k=2\pi/\lambda$ is the wave number, $J_q$ denotes the $q$-th order Bessel function of the first kind and $\chi(\xi)=\beta(\xi)+k\xi^2/2z$. The integral in Eq. (5) corresponds to the $q$-th order Hankel transform of the radial function $a(\xi)\exp[i\chi(\xi)]$.

Now, let us assume that we desire an AV with radius $r_0$. We determine the radial phase $\beta(\xi)$ for which this focal field is optimum. From Eq. (5) we can establish the AV intensity at $r=r_0$ as

$$|F(r_0)|^2 = \left|\int_0^\infty f_{pos}(\xi)\exp[i\chi(\xi)]\operatorname{sgn}\left\{J_q\left(2\pi\frac{r_0}{\lambda z}\xi\right)\right\}d\xi\right|^2, \quad (6)$$

where $f_{pos}(\xi)=\xi a(\xi)|J_q(2\pi r_0\xi/\lambda z)|$ is a non-negative real function, and 'sgn{x}' is a binary function, equal to +1 for x≥0, and −1 otherwise. Since the integrand in Eq. (6) is formed by the non-negative function $f_{pos}(\xi)$ multiplied by phase factors, we can apply the continuous form of the triangle inequality [18] to obtain the relation

$$|F(r_0)|^2 \leq \left(\int_0^\infty f_{pos}(\xi)d\xi\right)^2, \quad (7)$$

where the squared integral represents the upper bound value for $|F(r_0)|^2$. It is straightforward to establish from Eq. (6) that the intensity $|F(r_0)|^2$ achieves this upper bound value if $\exp[i\beta(\xi)]=\exp(-ik\xi^2/2z)\operatorname{sgn}[J_q(2\pi r_0\xi/\lambda z)]$. Therefore, the phase modulation of the plate [Eq. (4)] that generates the optimum AV of radius $r_0$ is

$$p(\xi,\phi) = \exp\left(-i\frac{k\xi^2}{2z}\right)\exp[-i\alpha(\xi,\phi)] \\ \operatorname{sgn}\left\{J_q\left(2\pi\frac{r_0}{\lambda z}\xi\right)\right\}\exp(iq\phi). \quad (8)$$

The phase plate with the transmittance in Eq. (8), illuminated by the input beam $g(\xi,\phi)$ [Eq. (3)], transmits the field $f(\xi,\phi)=a(\xi)\exp(-ik\xi^2/2z) \operatorname{sgn}\{J_q(2\pi r_0\xi/\lambda z)\} \exp(iq\phi)$. Because of the quadratic phase factor in $f(\xi,\phi)$, the complex amplitude of the AV, at the distance $z$ from the plate, is equivalent to the Fourier transform of the function $a(\xi) \operatorname{sgn}\{J_q(2\pi r_0\xi/\lambda z)\} \exp(iq\phi)$.

An important input field that belongs to the class defined in Eq. (3) is the GB, whose complex amplitude can be expressed, omitting factors that are independent of $\xi$, as

$$g(\xi,\phi) = \exp(-\xi^2/w^2)\exp(ik\xi^2/2R), \quad (9)$$

where $w$ is the beam radius and $R$ is the curvature radius of the quadratic phase, at the plate plane. In order to apply the general results to the case of the input GB, we only have to replace the amplitude and phase in Eq. (3) by $\exp(-\xi^2/w^2)$, and $\exp(ik\xi^2/2R)$, respectively. Thus, the plate transmittance that transforms the input GB into an optimum AV, with topological charge $q$, is

$$p(\xi,\phi) = \exp\left\{-i\frac{k}{2}\left(\frac{1}{R}+\frac{1}{z}\right)\xi^2\right\} \\ \operatorname{sgn}\left\{J_q\left(2\pi\frac{r_0}{\lambda z}\xi\right)\right\}\exp(iq\phi). \quad (10)$$

Note that the quadratic phase factor in Eq. (10), corresponds to the transmittance of a conventional lens, which is responsible of the basic focusing. On the other hand, the annular form of the focal field, with optimum peak intensity, is allowed by the last two factors, in both Eq. (8) and Eq. (10), which correspond to the phase modulation of the $q$-th order Bessel beam of radial spatial frequency $\rho_0=r_0/\lambda z$. Two illustrative examples of the phase modulation in Eq. (10), with topological charges $q=0$ and $q=1$, respectively, are depicted in Fig. 2.

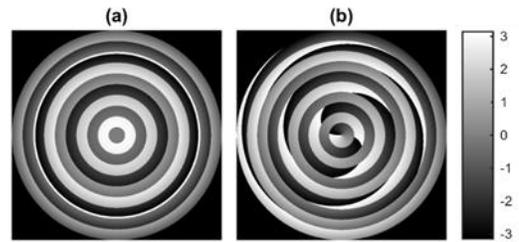

Fig. 2. Central sections in phase modulations of the phase plates that generate optimal annular vortices of topological charges (a) q=0, and (b) q=1, employing an input Gaussian beam.

Our discussion and results are connected with previous works dealing with the so called perfect AV [13-16] which is an infinitely narrow annular focal field with arbitrary integer topological charge. It is clear that this ideal field cannot be generated in practice. However, the optimum physically realizable approximation to the perfect AV, employing the optical setup in Fig. 1, is generated by the phase plate whose transmittance is given by Eq. (8), in the general case, or Eq. (10), for an input GB. Such phase transmittances can be, in principle, fabricated by lithography on a glass substrate. An attractive option, discussed below, is the use of a phase liquid-crystal (LC) spatial light modulator (SLM).

For the generation of an optimum AV, with the described approach, we employ the experimental setup depicted in Fig. 2. In this setup, the light beam from a He–Ne laser source (LS) is cleaned and expanded by a spatial filter (SF), and collimated by a lens (L). The result is a GB that illuminates a reflecting phase SLM (Model PLUTO, Holoeye Photonics LG). In the SLM we implemented the phase for optimum annular focusing of the GB [Eq. (10)] with an additional one-dimensional linear phase modulation, whose purpose is to separate the focal field from the un-modulated fraction of the light reflected by the SLM. The AV is recorded by a CCD, which is not shown at the setup.

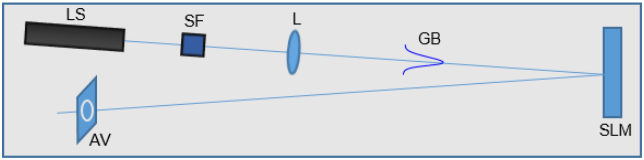

Fig. 3. Optical setup to generate the optimum annular focusing of a GB using a phase SLM.

In the experiment, we employed the parameter $\rho_0^{-1}=\lambda z/r_0 =208$ μm and estimated that the radius of the generated Gaussian beam, at the SLM plane, was $w \cong 5\rho_0^{-1}$, while the curvature radius of the quadratic phase was $R \cong 10$m. The other parameters employed in the design of the phase filter $p(\xi,\phi)$ are $\lambda=632.8$ nm, $z=50$ cm, and the topological charges $q=0, 4$. With these parameters, the expected diameter of the AVs is approximated to 3 mm. Fig. 4 (a, c) show the intensity of the AV (recorded by the CCD) and its transverse intensity profile for the case $q=0$. The result for $q=4$, which is not displayed, is almost identical to the one for $q=0$. For comparison with the experimental result, in Fig. 4 (b,d) we display the AV and its normalized intensity profile, which were numerically computed with Eq. (5) considering the beam radius $w=5\rho_0^{-1}$ and the other parameters employed in the experiment.

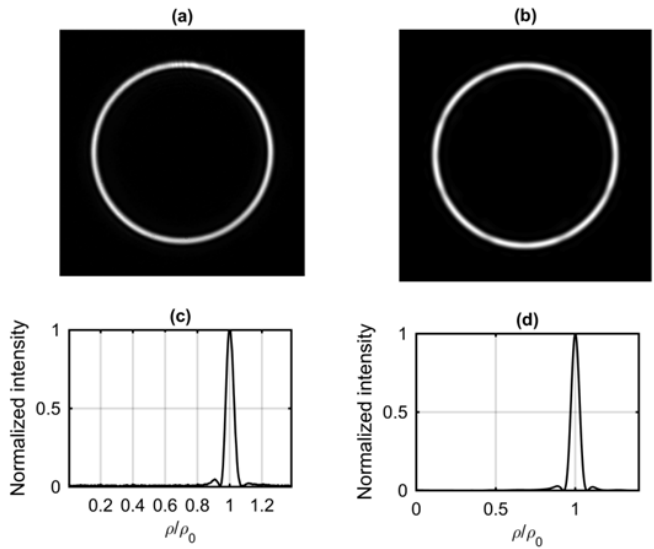

Fig. 4. (a) Experimentally recorded intensity, and (c) transverse intensity profile of an optimum AV with topological charge q=0, generated by a GB of radius $w \cong 5\rho_0^{-1}$. The numerically simulated results, with similar parameters, are shown in (b) and (d).

The relative annulus width in the AV is controlled by the radius of the Gaussian beam. E. g. the normalized AV intensity profiles computed when $w$ takes the values $3\rho_0^{-1}$ and $10\rho_0^{-1}$ (and $q=0$) are displayed in Fig. 5. It is clearly noted the inverse dependence between the GB radius $w$ and the AV transverse width.

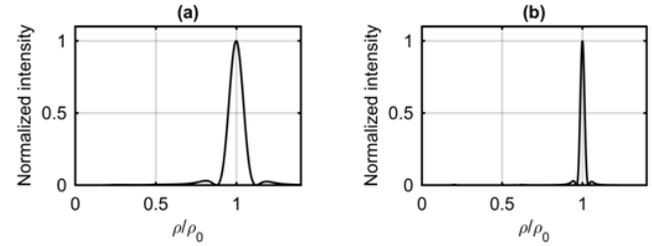

Fig. 5. Transverse intensity profiles of optimal AVs with topological charge q=0, generated by a GB of radius (a) $w=3\rho_0^{-1}$ and (a) $w=10\rho_0^{-1}$.

In other experiments and numerical simulations we have obtained that the peak intensities and widths of the optimal AVs (generated by GBs) show very low variation when $w$ is fixed and the topological charge $q$ takes different values. We computed the peak intensity and the full width at half of the maximum intensity (FWHM) in AVs obtained for different values of $w$ and $q$. The results, shown in Fig. 6, correspond to $w=5\rho_0^{-1}$ (left) and $w=10\rho_0^{-1}$ (right). The peak intensities for different $q$'s and a fixed $w$ are normalized respect to the peak intensity for $q=0$, and the FWHM's (in spatial frequency units) are normalized respect to $\rho_0$. In general we have found that for $w=Q\rho_0^{-1}$, the peak intensities (and corresponding FWHM's) of optimum AVs, present a relatively low variance for topological charges $q$ in the range $[0,2Q]$. The $q$-values for each plot in Fig. 6 correspond to this range.

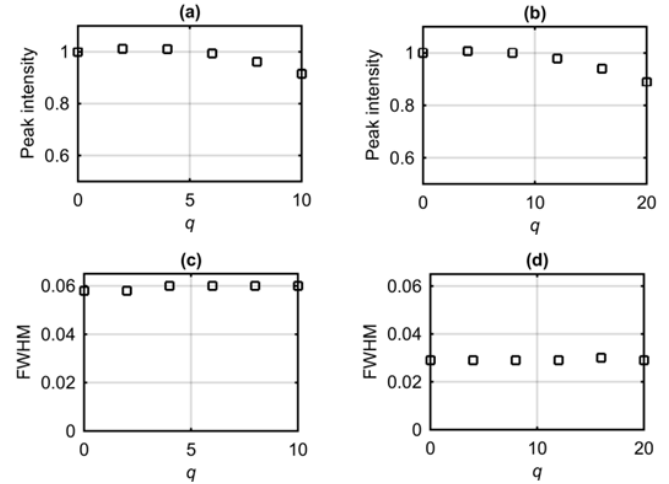

Fig. 6. Normalized peak intensities (top) and FWHMs (bottom) of optimal AVs for several topological charges $q$, generated by a GB of radius (a,c) $w=5\rho_0^{-1}$ and (b,d) $w=10\rho_0^{-1}$.

Summarizing, we have established the transmittance of a phase plate that allows the optimum focusing of a beam into an AV with arbitrary integer topological charge $q$. This result is an important complement of the well-known case of optimum focusing into a single spot. The annular focal field generated by the discussed phase plate corresponds to the optimum physically realizable implementation of the perfect AV, in the context of the employed setup. It is not difficult to understand that other desirable features of the optimum AV, as narrow annulus and high intensity gradient, are natural consequences of maximizing the peak intensity.

The phase modulation of the high order Bessel beams, that appear as main factor in the transmittance of the designed phase plates, has been previously employed in the efficient although approximate generation of Bessel-Gauss beams [19-21].

We illustrated by means of numerical simulations and experimentally the generation of optimal AVs generated by a GB. An interesting result is that the peak intensities and widths of the AVs generated with the proposed method, show very low variance if the width of the GB is fixed and the topological charge is modified in a certain range of values. The radius and relative transverse width (FWHM) of the optimal AVs are controlled by the parameters of the phase plate transmittance and the width of the input GB. Specifically, the annulus radius $r_0$ appears explicitly in the analytical expression for the plate transmittance.

The maximum peak intensity, large gradient and narrow annulus are advantageous attributes of the optimal AVs in different applications of these fields.

## REFERENCES

1. M. Born and E. Wolf, *Principles of optics*, (Cambridge University Press, 2003).
2. H. He, M. E. J. Friese, N. R. Heckenberg, and H. Rubinsztein-Dunlop, "Direct observation of transfer of angular momentum to absorptive particles from a laser beam with a phase singularity," Phys. Rev. Lett. **75**, 826–829 (1995).
3. V. Garces-Chavez, D. McGloin, M. J. Padgett, W. Dultz, H. Schmitzer and K. Dholakia, "Observation of the transfer of the local angular momentum density of a multiringed light beam to an optically trapped particle," Phys. Rev. Lett. **91**, 093602(2003).
4. A. M. Amaral, E. L. Falcão-Filho, and C. B. de Araújo, "Shaping optical beams with topological charge," Opt. Lett. **38**, 1579-1581 (2013).
5. M. Levenson, S. M. Tan, G. Dai, Y. Morikawa, N. Hayashi, and T. Ebihara, "The Vortex Via Process: Analysis and Mask Fabrication for Contact CDs < 80nm," Proc. SPIE 5040, 344-370 (2003).
6. Y. Liu, D. Liu and J. Hu, "Single Exposure General Vortex Phase-shift Mask for Contact Hole, " Proc. SPIE 5567, 723-731 (2004).
7. V. Westphal and S.W. Hell, "Nanoscale resolution in the focal plane of an optical microscope," Phys. Rev. Lett. **94**, 143903 (2005).
8. H. H. Arnaut and G. A. Barbosa, "Orbital and intrinsic angular momentum of single photons and entangled pairs of photons generated by parametric down-conversion," Phys. Rev. Lett. **85**, 286-289 (2000).
9. A. Mair, A. Vaziri, G. Weihs, and A. Zeilinger, "Entanglement of the orbital angular momentum states of photons," Nature (London) **412**, 313-316 (2001).
10. S. Franke-Arnold, S. M. Barnett, M. J. Padgett, and L. Allen, "Two-photon entanglement of orbital angular momentum states," Phys. Rev. A **65**, 033823 (2002).
11. D. Mawet, P. Riaud, O. Absil, and J. Surdej, "Annular groove phase mask coronagraph, " Astrophys. J. **633**, 1191-1200 (2005).
12. J. H. Lee, G. Foo, E. G. Johnson, and G. A. Swartzlander Jr., "Experimental verification of an optical vortex coronagraph," Physical review letters **97**, 053901 (2006).
13. A. S. Ostrovsky, C. Rickenstorff-Parrao, and V. Arrizón, "Generation of the 'perfect' optical vortex using a liquid-crystal spatial light modulator," Opt. Lett. **38**, 534–536 (2013).
14. M. Chen, M. Mazilu, Y. Arita, E. M. Wright, and K. Dholakia, "Dynamics of microparticles trapped in a perfect vortex beam," Opt. Lett. **38**, 4919–4922 (2013).
15. J. García-García, C. Rickenstorff-Parrao, R. Ramos-García, V. Arrizón, and A. S. Ostrovsky, "Simple technique for generating the perfect optical vortex," Opt. Lett. **39**, 5305-5308 (2014).
16. P. Vaity and L. Rusch, "Perfect vortex beam: Fourier transformation of a Bessel beam," Opt. Lett. **40**, 597-600 (2015).
17. V. Arrizón, U. Ruiz, D. Sánchez-de-la-Llave, G. Mellado-Villaseñor, and A. S. Ostrovsky, "Optimum generation of annular vortices using phase diffractive optical elements," Opt. Lett. **40**, 1173-1176 (2015).
18. A. David Wunsch, *Complex Variables with Applications* (Addison-Wesley Publishing Company, Inc., 1994).
19. V. Arrizón, D. Sánchez-de-la-Llave, U. Ruiz, and G. Méndez, "Efficient generation of an arbitrary nondiffracting Bessel beam employing its phase modulation," Opt. Lett. **34**, 1456–1458 (2009).
20. M. McLaren, J. Romero, M. J. Padgett, F. S. Roux, and A. Forbes, "Two-photon optics of Bessel-Gaussian modes," Phys. Rev. A **88**, 033818 (2013).
21. V. Arrizón, U. Ruiz, D. Aguirre-Olivas, D. Sánchez-de-la-Llave, and A. S. Ostrovsky, "Comparing efficiency and accuracy of the kinoform and the helical axicon as Bessel–Gauss beam generators," J. Opt. Soc. Am A **31**, 487-492 (2014).